\documentclass[aps,prl,twocolumn,groupedaddress,longbibliography]{revtex4-2}

\usepackage{graphicx}
\usepackage{dcolumn}
\usepackage{bm}
\usepackage{color}

\begin{document}

\title{Nonlinear periodic and solitary rolling waves in falling two-layer viscous liquid films}

\author{Andrey Pototsky}
\email[]{apototskyy@swin.edu.au}
\affiliation{Department of Mathematics, Swinburne University of Technology, Hawthorn, VIC 3122, Australia}

\author{Ivan S.~Maksymov}
\affiliation{Optical Sciences Centre, Swinburne University of Technology, Hawthorn, VIC 3122, Australia}
\affiliation{Artificial Intelligence and Cyber Futures Institute, Charles Sturt University, Bathurst, NSW 2795, Australia}

\date{\today}

\begin{abstract}
We investigate nonlinear periodic and solitary two-dimensional rolling waves in a falling two-layer liquid film in the regime of non-zero Reynolds numbers. At any flow rate, a falling two-layer liquid film is known to be linearly unstable with respect to long-wave deformations of the liquid-air surface and liquid-liquid interface. Two different types of zero-amplitude neutrally stable waves propagate downstream without growing or shrinking: a zig-zag surface mode and a thinning varicose interface mode. Using a boundary-layer reduction of the Navier-Stokes equation,  we investigate the onset, possible bifurcations and interactions of nonlinear periodic travelling waves. Periodic waves are obtained by continuation as stationary periodic solutions in the co-moving reference frame starting from small-amplitude neutrally stable waves. We find a variety of solitary waves that appear when a periodic solution approaches a homoclinic loop. Wave interactions are studied using direct numerical simulations of the boundary-layer model. We reveal that in the early stages of temporal evolution coarsening is dominated by an inelastic collision and merging of waves that travel at different speeds. Eventually, coarsening becomes arrested when the waves have reached the largest admissible amplitude. Bifurcation analysis confirms the existence of an upper bound of possible solitary wave amplitudes, thus explaining the arrest of coarsening. In the mixed regime, when both mode types are unstable, the temporal dynamics becomes highly irregular due to the competition between a faster-travelling zig-zag mode and a slower-travelling varicose mode. A quintessentially two-layer dynamical regime is found, which corresponds to a ruptured second layer. In this regime, the first layer adjacent to the solid wall acts as a conveyor belt, transporting isolated rolling droplets made of the second fluid downstream.
\end{abstract}


\maketitle

\setcounter{secnumdepth}{1}
\section{Introduction}
\label{intro}
The pioneering experiments by Kapitza \& Kapitza \cite{Kapitza49} on the onset and formation of nonlinear surface waves in a viscous liquid layer flowing down an inclined plane have set the stage for an avalanche of theoretical studies on hydrodynamic stability and nonlinear dynamics of the laminar flow in liquid layers in the presence of a deformable surface. For a single layer with a flat surface flowing down an inclined plane with a no-slip boundary, the base flow, also known as the Nusselt flow, has a parabolic longitudinal velocity profile with the maximum velocity attained at the free surface. The linearization of the Navier-Stokes equation about the base flow predicts the onset of a long-wavelength instability at a critical Reynolds number ${\rm Re}_c=5/4\cot{\theta}$, where $\theta$ is the inclination angle and ${\rm Re}=u_0d/\nu$ is the Reynolds number associated with the Nusselt flow with a characteristic speed $u_0$ in the film of thickness $d$ and kinematic viscosity $\nu$ \cite{benjamin57, Yih63}. Thus, for a falling liquid layer, i.e. when $\theta=\pi/2$, the base flow is unstable at any flow rate, no matter how small. As a result of the instability, small amplitude surface waves grow (decay) if their wavelength is larger (smaller) than a certain critical wavelength, which decreases with increasing flow rate. For falling liquid films, the neutrally stable waves with the critical wavelength propagate downstream with a velocity independent of their wavelength and exactly equal to twice the maximum Nusselt flow velocity.  As the instability develops, both amplitude and wavelength of the surface waves increase leading to the formation of a wave train of periodic nonlinear waves, chaotic waves, or a single solitary wave whose wavelength exceeds its amplitude or the average thickness of the carrier liquid layer by orders of magnitude \cite{portalski72,alekseenko85,trifonov_tsvelodub_1991,Gollub94,Dukler95,chang1993,CHANG96}. It was also observed that the travelling speed of nonlinear waves can be either larger or smaller than the speed of the neutrally stable small amplitude waves. Usually, the waves that propagate faster than the neutral waves are associated with drops or bulges, while those that propagate slower than the neutral waves can be identified with regions of large amplitude depressions, or holes.

The effort to describe the dynamics of nonlinear surface waves has primarily concentrated on considering small to moderate Reynolds numbers while assuming long-wave deformations of the free surface. Several competing and (or) complementary models have been derived, including the Benney model \cite{Benney66} that uses a systematic expansion of the hydrodynamic equations for long-wave disturbances of the base flow at finite Reynolds numbers, the Kuramoto-Sivashinski equation \cite{Sivash} that employs a weakly nonlinear reduction of the Benney model, the zero-Reynolds number long-wave model \cite{Oron97} that is derived from a Stokes equation in the limit of long-waves deformations of the film surface, and the boundary-layer model developed by Shkadov \cite{Shkadov67,Shkadov68}, where a self-similar parabolic longitudinal velocity profile was assumed and the validity of such idealisation confirmed at finite Reynolds numbers. Extensive numerical and theoretical analysis of these and other relevant models revealed a high complexity of the nonlinear waves dynamics and their interaction. For a complete overview of the wave dynamics in one-layer liquid films on an inclined plane, we refer the reader to several original works \cite{alekseenko85,trifonov_tsvelodub_1991,Nguyen2000,Gollub94,Chang89,CHANG87,pumir83,thiele04,chang1993} as well as several comprehensive reviews \cite{CHANG96,Craster09,kalliadasis2011falling}.

Any nonlinear propagating surface wave in a one-layer film can be described in the co-moving frame of reference as a stationary periodic orbit of an effective three-dimensional dynamical system. From the point of view of the dynamic system theory, a solitary wave is formed, when the period of the periodic orbit diverges. This occurs as a result of the Shilnikov homoclinic bifurcation, i.e. when the periodic orbit collides with a hyperbolic fixed point with one real and a pair of two complex conjugate eigenvalues \cite{Shilnikov,CHANG87}.  In real physical systems with finite lateral extent, a solitary wave can be formed due to coarsening from unstable periodic waves. Generally, two coarsening modes can be identified: one mode (the translational mode) is related to collision between two waves travelling at different speeds, and the other mode (mass transfer mode, or Oswald ripening mode) is related to the convective mass transfer from one wave into another without collision \cite{thiele_coars}.

Significantly less attention has thus far been given to hydrodynamic instabilities in multilayered liquid films with deformable interfaces flowing down an incline. Earlier theoretical works focused on the linear stability of the base flow in two-layer films composed of two immiscible fluids with a free upper surface flowing down an inclined plane \cite{Kao65,kao1968,Lowenherz89}. In the limit of long wave perturbations, two different instability modes have been identified: the first mode (interfacial mode) corresponds to a wave formed predominantly at the interface between two fluids, while the second mode (surface mode) represents the waves formed predominantly at the upper surface. It was also shown that the surface waves are always faster than the interfacial waves \cite{Kao65,kao1968}.
   Nonlinear surface waves in multi-layered liquid films on an incline have been studied in Ref.\,\cite{kliakhandler1999}, where a set of coupled Kuramoto-Sivashinsky equations has been derived under the assumption of long-wave deformations of the interfaces. Upon taking into account the inertia effects, the long-wave instability of a vertically falling two-layer film, studied earlier in \cite{Kao65,kao1968} was recovered and the nonlinear temporal evolution of the interfaces was exemplified by means of numerical simulations. Periodic solutions in linearly coupled Kuramoto-Sivashinsky-Korteweg-de Vries equations were investigated in Refs.\,\cite{Malomed02}, where it was demonstrated that some solitary wave solutions associated with cnoidal waves may be linearly stable, including two-dimensional spatially localized pulses \cite{Malomed02a}. In Ref.~\cite{Kliakhandler98} the nonlinear theory of long Marangoni waves in horizontal two-layer liquid films has been developed based on the amplitude equation derived from the Navier-Stokes equation in the weakly nonlinear regime. Travelling surface waves have been found to form as a result of oscillatory Marangoni instability and spontaneous spatial symmetry breaking. The combined effect of the horizontal temperature gradient and gravity on the onset and formation of nonlinear wavy patterns in ultra-thin horizontal two-layer films at zero Reynolds number has been investigated in Ref.\, \cite{nepomnyashchy2010}. Numerical simulations revealed a rich variety of two-dimensional travelling waves, including propagating fronts, stripes, holes and droplets. Nonlinear dynamics of inertialess two-layer film on an incline was studied using direct numerical simulations of the Navier-Stokes equations at zero Reynolds numbers in Ref.~\cite{Jiang04}. 

 Probably the most comprehensive theoretical and experimental study of a two-layer flow on an inclined corrugated substrate at zero Reynolds number was reported recently in Ref.\cite{shah21}. Assuming the Stokes flow limit and taking into account long-wave deformations, a reduced hydrodynamic model was derived that can be directly compared with earlier models of inertialess two-layer films \cite{pot05}. The model was used to study the dynamics of propagating flow fronts sliding down an inclined corrugated plane in the presence of contact lines. Similarity solutions were constructed to examine possible dynamic regimes. It was shown experimentally and theoretically that a small amount of a heavier fluid can accelerate an enveloping upper layer of a lighter fluid propagating ahead of the heavier layer \cite{shah21}.

Here we use a two-layer version of the Shkadov model \cite{bespot16,pot19} that was developed earlier to study nonlinear periodic and solitary waves in a free falling two-layer liquid film with a free upper surface in the regime of finite Reynolds numbers.
We demonstrate that the speed of the neutrally stable waves predicted by the integrated boundary layer model exactly coincides with the earlier result obtained from the full Navier-Stokes equation linearized about the base flow in the limit of long-wave perturbations \cite{Kao65,kao1968}. After validating the model, we study nonlinear periodic waves that bifurcate from the small-amplitude neutrally stable waves and can be found using the numerical continuation method \cite{krauskopf,dijkstra2014,doedel1981auto} as stationary travelling wave solutions in the co-moving frame of reference. Solitary waves are obtained as a result of a homoclinic bifurcation of periodic solutions, i.e.~when the period (wavelength) diverges. Depending on the ratio of the fluid viscosities, densities, surface tensions and film thicknesses, we find a variety of different solitary waves, characterized by the absolute and relative deflections of the interface and the upper surface.
We present a detailed bifurcation analysis of the periodic solutions and make a connection with the Shilnikov theorem for saddle-focus homoclinics in higher dimensions \cite{Shilnikov2}.
Direct numerical simulations of the dynamic equations reveal that in the early stages of the temporal evolution from an initially flat film, coarsening is dominated by the translational mode, whereby the faster-travelling larger waves collide and merge with the slower-travelling smaller waves. Bifurcation analysis of periodic waves predicts the existence of the solitary wave with the largest possible amplitude for any give combination of the average film thicknesses.  As the result, in the systems with conserved volume, coarsening is completely arrested in the late stages of temporal evolution, when the waves grow to reach their maximum admissible amplitude. This observation is in agreement with earlier studies of rolling waves in the shallow-water equation \cite{balmforth_mandre_2004} and in granular matter \cite{Razis14}.

The results presented in continuation may find practical applications in the areas, where the interactions between two fluid layers play important roles, which include but not limited to glaciology \cite{Kow20}, volcanology \cite{Gri00}, hydrology \cite{Woo00} and biology \cite{Ram20}. Moreover, our findings also generalize the well-established, in the case of single falling liquid films \cite{Gollub94, Ker94}, fact that the ability of solitary waves in falling liquid films to merge sharply contrasts with the collision behaviour of solitary waves observed in other physical \cite{Rem_book, Kivshar_book} and biological systems \cite{Hei05, Gon14} (see Sec.~\ref{sec2} ). Apart from implications for fundamental nonlinear physics, this particular behaviour of liquid-film solitary waves may be of interest in the emergent sub-field of nonlinear dynamics concerned with the development of physical analogues of artificial neural networks \cite{RC_book}, where it has been demonstrated that interactions between solitons can be used to mimic the operation of certain digital machine learning algorithms \cite{Sil21}.

\section{Long-wave instability of a falling two-layer film with inertia}
\label{sec1}
\begin{figure}[ht]
   \includegraphics[width=\columnwidth]{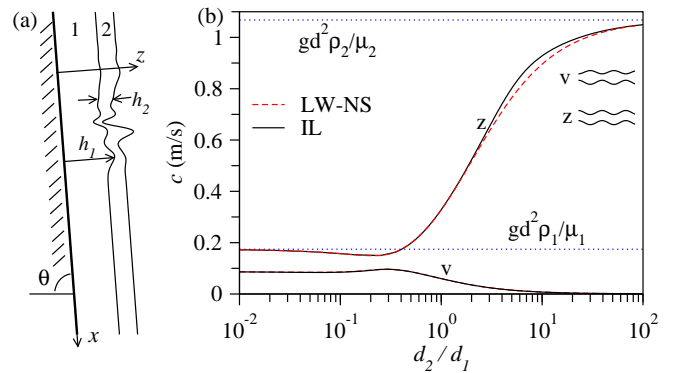}
   \caption{\label{F1}(a) A two-layer film flowing down an inclined plane. (b) Speed $c$ of neutrally stable waves {\it vs} film thickness ratio $d_2/d_1$ for a two-layer film with total thickness $0.5$ mm composed of oil as fluid (1) with  $\mu_1=0.026$~Pa\,s and $\rho_1=1850$~kg/m$^3$ and isopropanol as fluid (2) with $\mu_2=0.0018$~Pa\,s and $\rho_2=785$ (kg/m$^3$). The respective surface tensions are $\sigma_1=0.0063$~N/m and $\sigma_2=0.024$~N/m. Faster waves are of the zig-zag type (z), and slower waves are of the varicose type (v). Solid lines correspond to the numerical solution of the integrated-layer model Eq.\,(\ref{lin_eq4}). Dashed lines correspond to the analytical result from the full Navier-Stokes equation in the long-wave limit Eq.\,(\ref{eq_c})  and Ref.\,\cite{Kao65}.
   }
   \end{figure}

We consider a two-dimensional multilayer liquid film composed of two immiscible and incompressible fluids with dynamic viscosities $\mu_i$, densities $\rho_i$ and the unperturbed thicknesses $d_i$, $(i=1,2)$, with the fluid $i=1$ resting on a solid plate inclined at angle $\theta$ with respect to the horizon, as shown in Fig.\ref{F1}(a). The surface tensions of the liquid-liquid and the  liquid-air interfaces are $\sigma_1$ and $\sigma_2$, respectively. The longitudinal flow velocities $u_\parallel^{(i)}(x,z,t)$ in each layer are assumed to be quadratic functions of the vertical distance from the solid plate $z$. In the long-wave approximation, the Navier-Stokes equation in each fluid can be then integrated over $z$ to yield the first-order equations for the longitudinal fluxes $q_1=\int_0^{h_1}u_\parallel^{(1)}(z,x,t)\,dz$ and $q_2=\int_{h_1}^{h_1+h_2}u_\parallel^{(2)}(z,x,t)\,dz$  in the functional form \cite{bespot16,pot19}
\begin{eqnarray}
  \label{eq1}
  \rho_1\left(\partial_t q_1 + \partial_x({\bm q}^{\sf T} \cdot {\bm b}_1 \cdot {\bm q})\right)&=&
  \mu_1({\bm a}\cdot {\bm q})_1-h_1\partial_x \left(\frac{\delta F}{\delta h_1}\right),\nonumber\\
  \rho_2\left(\partial_t q_2 + \partial_x({\bm q}^{\sf T} \cdot {\bm b}_2  \cdot {\bm q})\right)&=&
  \mu_1({\bm a}\cdot {\bm q})_2-h_2\partial_x \left(\frac{\delta F}{\delta h_2}\right)\nonumber\\
\end{eqnarray}
where  ${\bm q}=(q_1;q_2)$ is a column $(2\times 1)$ vector of fluxes, and $(\dots)^T$ and $(\cdot)$ stand for the transposition and the scalar product, respectively.  The energy functional $F$ is given by
\begin{eqnarray}
\label{energy}
F&=&\int \left[ g \cos{\theta} \left( \frac{\rho_1  h_1^2}{2} +\rho_2h_1 h_2 +
\frac{\rho_2h_2^2}{2} \right)  \right.\nonumber\\
&-&\left. g\sin{\theta} (\rho_1 h_1 +\rho_2h_2)x \right.\nonumber\\
&+&\left. \frac{\sigma_1(\partial_x h_1)^2}{2}+
\frac{\sigma_2(\partial_x h_1+\partial_x h_2)^2}{2}\right]\,dx.
\end{eqnarray}
The functional derivatives in Eq.\,(\ref{eq1}) can be found explicitly as
\begin{eqnarray}
  \label{eq1a}
  \frac{\delta F}{\delta h_1}&=&-(\sigma_1+\sigma_2 )\partial_x^2 h_1 -\sigma_2 \partial_x^2 h_2 + g \cos{\theta} (\rho_1 h_1+\rho_2 h_2)\nonumber\\
  & -&\rho_1 g x \sin(\theta) ,\nonumber\\
   \frac{\delta F}{\delta h_2}&=&-\sigma_2 \partial_x^2 (h_1+h_2) + g \rho_2 \cos{\theta} (h_1+h_2) -\rho_2 g x \sin(\theta)\nonumber\\
\end{eqnarray}
In Eq.\,(\ref{eq1}) ${\bm b}_i={\bm m}^{\sf T} \cdot {\bm e}_i\cdot {\bm m}$ and matrices ${\bm a}$, ${\bm m}$ and ${\bm e}_i$ are given by
\begin{eqnarray}
  \label{matrices}
        {\bm a}&=&D^{-1}\left(\begin{array}{cc}
          -2h_1h_2^2(\frac13 \mu h_2+h_1) & h_1^3h_2\\
          h_1^2h_2^2 & -\frac23 h_1^3h_2
        \end{array}\right),\nonumber\\
         {\bm m}&=&D^{-1}\left(\begin{array}{cc}
          -h_2^2(\frac13 \mu h_2 +h_1) & \frac12 h_1^2h_2\\
          h_1^2h_2 & -\frac23h_1^3
        \end{array}\right)\nonumber\\
   {\bm e}_2&=&\left(\begin{array}{cc}
          h_1^4 h_2 & h_1^2 h_2^2\left(h_1+\frac{1}{3}\mu h_2\right)\\
         h_1^2 h_2^2\left(h_1+\frac{1}{3}\mu h_2\right) & \frac{2}{3}h_2^3\left(\frac{1}{5}\mu^2h_2^2+\mu h_1 h_2 +\frac{3}{2}h_1^2\right)
         \end{array}\right),\nonumber\\
         {\bm e}_1&=&\frac13 h_1^3\left(\begin{array}{cc}
          \frac85 h_1^2 & \frac54 h_1h_2\\
         \frac54 h_1h_2 & h_2^2
        \end{array}\right),
\end{eqnarray}
with $D=2\mu h_1^3h_2^3/9+h_1^4h_2^2/6$ and $\mu=\mu_1/\mu_2$.  Note that matrices ${\bm e}_i$ and ${\bm b}_i$ are symmetric.

The system Eq.\,(\ref{eq1}) is complemented by two kinematic equations that represent the conservation of the volume of each layer
\begin{eqnarray}
\label{eq2}
\partial_t h_i +\partial_x q_i=0.
\end{eqnarray}

The system Eqs.\,(\ref{eq1},\ref{eq2}) represents the long-wave boundary-layer model of a two-dimensional two-layer viscous film.
By setting $\rho_1=\rho_2$, $\mu_1=\mu_2$, $\sigma_1=0$ and letting $q_1,h_1\rightarrow 0$, it can be shown that system Eqs.\,(\ref{eq1},\ref{eq2}) exactly reduces to Shkadov's model of a one-layer liquid film on an incline \cite{Shkadov67,Shkadov68}.

 Earlier Eqs.\,(\ref{eq1},\ref{eq2}) have been used to study Faraday instability in horizontal two-layer films \cite{potbes16,bespot16} and the dynamics of an isolated liquid drop of a heavier fluid on top of an immiscible lighter fluid \cite{pot19}. As it was shown in Ref.\,\cite{pot19}, in the limit of zero Reynolds number, Eqs.\,(\ref{eq1},\ref{eq2}) reduce to the inertialess long wave model of two-layer films \cite{pot05} in the parametrization used in Ref.\,\cite{Boomer}.

 The Shkadov model was shown to correctly predict the velocity of two-dimensional nonlinear surface waves in one-layer falling liquid films \cite{alekseenko85,trifonov_tsvelodub_1991}. It should be noted, however, that the two-dimensional flow regime can only be sustained for relatively small Reynolds numbers, ${\rm Re}=gd_2^3/[3(\mu_2/\rho_2)^2]\approx 1\dots 10$. As a result of this limitation, the thickness of the layers must be chosen in the sub-millimetre range for common fluids, such as water or oil.  The Shkadov model and its minor modifications have also been previously used to study Faraday instability in one-layer liquid films \cite{bes13,bes17} and in isolated liquid drops \cite{maksymov19}.

In what follows we consider a free-falling two-layer film on a vertical solid plate, i.e.~$\theta=\pi/2$.  The base stationary flow with constant fluxes $q_i^{(0)}$ corresponds to a flat two-layer film with $h_i=d_i$ and can be found from Eq.\,(\ref{eq1})
 \begin{eqnarray}
 \label{fluxes}
 \mu_1({\bm a}^{(0)}\cdot {\bm q}^{(0)})_i+\rho_i g d_i=0
 \end{eqnarray}
 where ${\bm a}^{(0)}$ represents the matrix ${\bm a}$ with $h_i=d_i$. Inverting the matrix ${\bm a}$, we obtain
 \begin{eqnarray}
 \label{stat_flux}
 q_1^{(0)}&=&\frac{g\rho_1}{2\mu_1} \left(\frac{2}{3}d_1^3+\frac{\rho_2}{\rho_1}d_1^2d_2 \right)\nonumber\\
  q_2^{(0)}&=&\frac{g\rho_1}{2\mu_1}\left(d_1^2 d_2+\frac{2\rho_2}{\rho_1}\left(d_2^3\frac{\mu_1}{3\mu_2}+d_1d_2^2\right)\right).
 \end{eqnarray}
 Stationary fluxes Eqs.\,(\ref{stat_flux}) exactly coincide with the steady Nusselt flow in a free-falling two-layer flat film \cite{Kao65,kao1968}.

The linear stability of the base flow is studied using a conventional plane wave ansatz, i.e. by setting $q_i=q_i^{(0)}+\tilde{q}_ie^{\lambda t +ikx}$ and $h_i=d_i+\tilde{h}_ie^{\lambda t +ikx}$, where $\lambda$ and $k$ are the perturbation growth rate and the wave vector, respectively.
 Differentiating Eq.\,(\ref{eq2}) with respect to time and Eq.\,(\ref{eq1}) with respect to $x$, and eliminating the mixed derivatives $\partial_{tx} q_i$ as well as replacing $\partial_x q_i$ with $-\partial_t h_i$, the linearized system can be reduced after some lengthy but straight-forward calculations to two coupled linear equations for the film thickness perturbations
 \begin{eqnarray}
 \label{lin_eq1}
 \left(
 \begin{array}{cc}
 \lambda^2 +A_{11}\lambda+B_{11} & A_{12}\lambda +B_{12}\\
  A_{21}\lambda +B_{22} & \lambda^2 +A_{22}\lambda+B_{22}
 \end{array}
 \right)
 \left(
 \begin{array}{c}
 \tilde{h}_1\\
 \tilde{h}_2
 \end{array}
 \right)=0,
 \end{eqnarray}
 where the matrices ${\bm A}$ and ${\bm B}$ are given by
  \begin{eqnarray}
 \label{lin_eq2}
 {\bm A}&=&-\frac{\mu_1}{\rho_1}
 \left(
 \begin{array}{cc}
  a_{11}^{(0)}& a_{12}^{(0)}\\
 \rho a_{21}^{(0)}& \rho a_{22}^{(0)}\\
 \end{array}
 \right)\nonumber\\
 &+&
 2ik
 \left(
 \begin{array}{cc}
 ({\bm b}_1^{(0)}\cdot {\bm q}^{(0)})_1 & ({\bm b}_1^{(0)}\cdot {\bm q}^{(0)})_2 \\
 ({\bm b}_2^{(0)}\cdot {\bm q}^{(0)})_1 & ({\bm b}_2^{(0)}\cdot {\bm q}^{(0)})_2 \\
 \end{array}
 \right)
 \end{eqnarray}
 and
 \begin{eqnarray}
 \label{lin_eq3}
 {\bm B}&=&
 k^2\left(
\begin{array}{cc}
  ({\bm q}^{(0)})^T\cdot \partial_{d_1}{\bm b}_1^{(0)}\cdot {\bm q}^{(0)}, & ({\bm q}^{(0)})^T\cdot \partial_{d_2}{\bm b}_1^{(0)}\cdot {\bm q}^{(0)} \\
  ({\bm q}^{(0)})^T\cdot \partial_{d_1}{\bm b}_2^{(0)}\cdot {\bm q}^{(0)}, & ({\bm q}^{(0)})^T\cdot \partial_{d_2}{\bm b}_2^{(0)}\cdot {\bm q}^{(0)}
 \end{array}
 \right)\nonumber\\
 \nonumber\\
 &+&\frac{ik\mu_1}{\rho_1}
 \left(
\begin{array}{cc}
(\partial_{d_1} {\bm a}^{(0)}\cdot {\bm q}^{(0)})_1, & (\partial_{d_2} {\bm a}^{(0)}\cdot {\bm q}^{(0)})_1\\
\rho (\partial_{d_1} {\bm a}^{(0)}\cdot {\bm q}^{(0)})_2, &\rho(\partial_{d_2} {\bm a}^{(0)}\cdot {\bm q}^{(0)})_2
\end{array}
 \right)\nonumber\\
 \nonumber\\
&+&\frac{k^4}{\rho_1}
 \left(
\begin{array}{cc}
d_1 (\sigma_1+\sigma_2) & d_1\sigma_2\\
\rho d_2 \sigma_2 & \rho d_2\sigma_2
\end{array}
\right)+ikg
 \left(
\begin{array}{cc}
1 & 0\\
0 & 1
\end{array}
\right),
 \end{eqnarray}
 where $\rho=\rho_1/\rho_2$ and superscript $(0)$ indicates that the corresponding matrices are obtained by replacing $h_i$ with $d_i$.

 The solvability condition for Eq.\,(\ref{lin_eq1}) is given by
 \begin{eqnarray}
 \label{lin_eq4}
&&( \lambda^2 +A_{11}\lambda+B_{11})( \lambda^2 +A_{22}\lambda+B_{22}) \nonumber\\
&=&(A_{12}\lambda +B_{12})(A_{21}\lambda +B_{22}).
  \end{eqnarray}
  Eq.\,(\ref{lin_eq4}) determines the dispersion relation $\lambda(k)$ and the ratio of (complex) deformation amplitudes
  \begin{eqnarray}
 \label{lin_eq5}
\frac{\tilde{h}_2}{\tilde{h}_1}&=&-\frac{ \lambda^2 +A_{11}\lambda+B_{11}}{A_{12}\lambda +B_{12}}
  \end{eqnarray}
Next, we simplify Eq.\,(\ref{lin_eq4}) in the leading order of the long-wave limit, i.e. when $k\rightarrow 0$ to obtain the analytical expression for the speed $c$ of neutrally stable waves with ${\Re}(\lambda)=0$.
  Thus, we anticipate $\lambda=-ikc$ and reduce Eq.\,(\ref{lin_eq4}) in the leading order $k\rightarrow 0$
     \begin{widetext}
  \begin{eqnarray}
 \label{lin_eq6}
 &&\left(\frac{k\mu_1}{\rho_1}\right)^2\left\{\left[ ca_{11}^{(0)} +(\partial_{d_1} {\bm a}^{(0)}\cdot {\bm q}^{(0)})_1+\frac{g\rho_1}{\mu_1}\right]\left[\rho c a_{22}^{(0)} +\rho (\partial_{d_2} {\bm a}^{(0)}\cdot {\bm q}^{(0)})_2 +\frac{g\rho_1}{\mu_1}\right]\right. \nonumber\\
 &-& \left.\left[ca_{12}^{(0)}  + (\partial_{d_2} {\bm a}^{(0)}\cdot {\bm q}^{(0)})_1\right]\left[\rho ca_{21}^{(0)}  + \rho (\partial_{d_1} {\bm a}^{(0)}\cdot {\bm q}^{(0)})_2 \right]\right\}=0 \,.
  \end{eqnarray}
   \end{widetext}

  Differentiating Eq.\,(\ref{fluxes}) with respect to $d_i$, we replace the derivatives $\partial_{d_i} {\bm a}^{(0)}$ in Eq.\,(\ref{lin_eq6}) to obtain

  \begin{eqnarray}
 \label{lin_eq7}
 &&\left[ ca_{11}^{(0)}  - ({\bm a}^{(0)}\cdot \partial_{d_1} {\bm q}^{(0)})_1 \right]\left[ ca_{22}^{(0)}  - ({\bm a}^{(0)}\cdot \partial_{d_2} {\bm q}^{(0)})_2 \right]\nonumber\\
 &=&\left[ ca_{12}^{(0)}  - ({\bm a}^{(0)}\cdot \partial_{d_2} {\bm q}^{(0)})_1 \right]\left[ ca_{21}^{(0)} - ({\bm a}^{(0)}\cdot \partial_{d_1} {\bm q}^{(0)})_2 \right]\nonumber\\
    \end{eqnarray}
    Solving the quadratic equation Eq.\,(\ref{lin_eq7}) we find after some algebra the speed $c$ of neutrally stable waves in the long-wave limit
    \begin{widetext}
    \begin{eqnarray}
 \label{lin_eq8}
 2c&=&(\partial_{d_2} q_2^{(0)}+\partial_{d_1}q_1^{(0)})\pm\sqrt{(\partial_{d_2} q_2^{(0)}+\partial_{d_1} q_1^{(0)})^2-4(\partial_{d_2} q_2^{(0)}\partial_{d_1} q_1^{(0)}-\partial_{d_1} q_2^{(0)}\partial_{d_2} q_1^{(0)})}.
    \end{eqnarray}
    \end{widetext}
   After substituting the derivatives of the fluxes $q_i^{(0)}$ from Eqs.\,(\ref{stat_flux}) into Eq.\,(\ref{lin_eq8}), we obtain the explicit expression for the speed of neutrally stable waves
    \begin{widetext}
     \begin{eqnarray}
 \label{eq_c}
 c&=&\frac{g}{4}\left(\frac{3 d_1^2 \rho_1}{\mu_1} + \frac{2 d_2^2 \rho_2}{\mu_2} + \frac{6 d_1 d_2\rho_2}{\mu_1} \pm \sqrt{\frac{d_1^4\rho_1^2}{\mu_1^2} + \frac{4d_1^3 d_2\rho_1\rho_2}{\mu_1^2} - \frac{4d_1^2d_2^2\rho_1\rho_2}{\mu_1\mu_2} + \frac{12d_1^2d_2^2\rho_2^2}{\mu_1^2} + \frac{8d_1d_2^3\rho_2^2}{\mu_1\mu_2} + \frac{4d_2^4\rho_2^2}{\mu_2^2}} \right).
    \end{eqnarray}
 \end{widetext}
 Eq.\,(\ref{eq_c}) exactly coincides with the speed of neutrally stable waves obtained earlier from the full Navier-Stokes equation as reported in Ref.\,\cite{Kao65} (note a few misprints in Ref.\,\cite{Kao65}: a missing factor of $1/2$ in front of the radical in Eq.(12) and a missing negative sign for $b_1$ in Eq.(1)).

As pointed out in Ref.\,\cite{Kao65}, the two signs in Eq.\,(\ref{eq_c}) correspond to two distinct neutral modes with one mode being always faster than the other. Eq.\,(\ref{lin_eq5}) can be used to show that the deflections of the liquid-liquid and the upper liquid-air interfaces are in-phase for the faster mode and out of phase for the slower mode. Therefore, according to the classification introduced in Ref.\cite{potbes16}, the faster (slower) mode can be associated with a zig-zag (varicose) wave, respectively. Another important property of the long neutral waves is that their speed  Eq.\,(\ref{eq_c})  does not depend on the surface tensions $\sigma_i$.

The accuracy of the long-wave expansion can be validated by comparing the speed of the neutral waves obtained from  Eq.\,(\ref{eq_c}) and from the nonlinear solvability condition Eq.\,(\ref{lin_eq4}). Thus we choose fluid parameters as used in the Faraday waves experiments in two-layer films \cite{Pucci11}. Fluid (1) is an oil layer with viscosity $\mu_1=0.026$~Pa\,s and density $\rho_1=1850$~kg/m$^3$ and fluid (2) is a significantly less viscous isopropanol layer with $\mu_2=0.0018$~Pa\,s and density $\rho_2=785$~kg/m$^3$.
  The total thickness of the two-layer film is fixed at $d=d_1+d_2=0.5$ mm to ensure that the characteristic Reynolds numbers are sufficiently small ${\rm Re}_1=gd^3/[3(\mu_1/\rho_1)^2]=2$ and ${\rm Re}_2=gd^3/[3(\mu_2/\rho_2)^2]=77.6$ to be able to sustain a two-dimensional flow regime.

In Fig.\ref{F1}(b) the speed of the neutral waves $c$ is shown as a function of the film thicknesses ratio $d_2/d_1$. The long-wave approximation Eq.\,(\ref{eq_c}) , which also coincides with the long-wave limit of the full Navier-Stokes equation \cite{Kao65} (dashed lines) is in excellent agreement with the numerical solution obtained from Eq.\,(\ref{lin_eq4}) (solid lines). In the limit of a vanishingly thin first (second) layer, the speed of the neutral waves approaches the corresponding one-layer limiting value that is given by either $gd^2 \rho_2/\mu_2$ for a vanishingly thin first layer or $gd^2 \rho_1/\mu_1$ for a vanishingly thin second layer. The faster wave is of the zig-zag type (z), while the slower wave is of the varicose type (v), as schematically shown in the inset of (b).

   %
   \section{Nonlinear periodic and solitary travelling waves}
   \label{sec2}
   %
   We are looking for periodic travelling waves in a two-layer liquid film that can be seen as periodic steady-state solutions in the co-moving frame of reference. By setting $h_i(x-vt)$ in Eq.\,(\ref{eq2}), where $v$ is the phase speed of the wave and integrating over $x$ we obtain a linear relationship between the fluxes $q_i$ and the local film thicknesses $h_i$
   \begin{eqnarray}
  \label{nonlin_eq1}
  -v h_i +q_i=f_i,
  \end{eqnarray}
  where $f_i$ are some constants of integration that depend on $v$.  It is important to emphasize that $f_i$ represent fluid fluxes in the co-moving frame of reference, where the wave profile remains time-independent.

    Then, we nondimensionalize Eqs.\,(\ref{eq1},\ref{eq2}) by scaling $h_i$ with the total film thickness $d=d_1+d_2$, fluxes $q_i$ with $g\rho_1 d^3/(3\mu_1)$, time $t$ with $\rho_1 d^2/\mu_1$ and $x$ with $g\rho_1^2 d^4/(3\mu_1^2)$ and the flow velocity with $g\rho_1 d^2/(3\mu_1)$. Using Eq.\,(\ref{nonlin_eq1}) and replacing $q_i(x,t)$ with $q_i(x-vt)$, we obtain from Eqs.\,(\ref{eq1},\ref{eq2})
      \begin{eqnarray}
  \label{nonlin_eq3}
  -v^2\partial_x h_1 + \partial_x({\bm q}^{\sf T} \cdot {\bm b}_1 \cdot {\bm q})&=&
  ({\bm a}\cdot {\bm q})_1+h_1 (\gamma_1+\gamma_2 )\partial_x^3 h_1  \nonumber\\
  &+& 3h_1+h_1\gamma_2 \partial_x^3 h_2,\nonumber\\
  -v^2\partial_x h_2 + \partial_x({\bm q}^{\sf T} \cdot {\bm b}_2  \cdot {\bm q})&=&
  \rho({\bm a}\cdot {\bm q})_2+3h_2\nonumber\\
  &+&\rho h_2 \gamma_2 \partial_x^3 (h_1+h_2),\nonumber\\
  q_i&=&f_i+vh_i.
\end{eqnarray}
where $\gamma_i=\sigma_i d/(\rho_1 (\mu_1/\rho_1)^2)$ and matrices ${\bm a}$ and ${\bm b}_i$ are the same as in Eqs.\,(\ref{matrices}).

Small-amplitude periodic waves bifurcate from the neutrally stable waves with speed $c$ and wavelength $k_c$ that both can be found from the dispersion relation $\lambda(k)$ by setting $\lambda(k_c)=-ick_c$.

Following standard methods for the travelling wave solutions \cite{thiele04}, we write Eqs.\,(\ref{nonlin_eq3}) as a boundary-value problem with periodic boundaries in the domain of length $L$. The initial domain length is a multiple of the neutrally stable wavelength, i.e. $L=2\pi n/k_c$, where $n$ is a positive integer. Solutions that correspond to a particular value of $n$ are called period-$n$ solutions. We use an efficient numerical continuation method described in \cite{krauskopf,dijkstra2014,doedel1981auto} with the starting point given by the analytically known small amplitude neutrally stable wave with speed $c$ and wave number $k_c$ that are found from Eq.\,(\ref{lin_eq4}). The boundary-value problem is supplemented with three integral conditions
  \begin{eqnarray}
  \label{nonlin_eq4}
  \int_{x=0}^L h_i(x)\,dx&=&Ld_i,~~\int_{x=0}^L h_1(x)\partial_x h_1^{old}(x)\,dx=0,\nonumber\\
  \end{eqnarray}
where $h_1^{old}(x)$ stands for the solution at the previous continuation step. The first two conditions in Eqs.\,(\ref{nonlin_eq4}) ensure that the average film thicknesses of the layers remain constant. The last condition in Eqs.\,(\ref{nonlin_eq4}) is necessary to fix the phase of the solution, due to the presence of translational symmetry. For any positive integer $n$, the neutral wave can be continued using $v$ as a principal continuation parameter. Thereby, the period of the wave $L$ and parameters $f_i$ from Eq.\,(\ref{nonlin_eq1}) are included in the list of continuation parameters and computed as a function of $v$ \cite{note}.

\begin{figure}[ht]
   \includegraphics[width=\columnwidth]{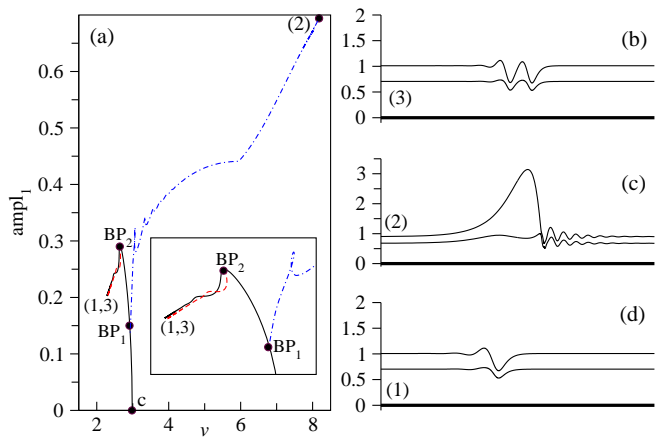}
   \caption{\label{F2}(a) Bifurcation diagram of periodic travelling waves for $\sigma=\gamma_i=\mu=\rho=1$ and $d_1/d=0.7$ The waves bifurcate from the zero-amplitude neutrally stable wave with speed $v=2.99$ (point c). Solutions are obtained by numerical continuation in parameter $v$ with three integral constraints given by Eqs.\,(\ref{nonlin_eq4}). The branch of period-one solutions (solid line) asymptotically approaches  the homoclinic bifurcation point $(1)$. Two new branches of period-two solutions bifurcate from the period-one branch at branching points BP$(1,2)$. The first period-two branch extends towards larger values of travelling speed and asymptotically approaches the homoclinic point $(2)$.  The second period-two branch asymptotically approaches the homoclinic point $(3)$.  (b,c,d) solitary wave solutions $(1,2,3)$
   }

   \end{figure}

As a reference case, we consider a fully symmetric system with $\gamma_i=\sigma=\mu=\rho=1$ and $d_1/d=0.7$. The faster zig-zag wave is linearly unstable for $k<k_c=1.6$, while the varicose wave remains linearly stable for all wave numbers. The bifurcation diagram in Fig.\,\ref{F2}(a) shows the amplitude ampl$_1={\rm max}(h_1)-{\rm min}(h_1)$ of the first layer in a periodic travelling wave as a function of $v$, where one can see that travelling waves bifurcate from the neutrally stable waves with the speed $c=2.99$. The branch of the period-one solutions obtained for $n=1$ is shown by the solid line. The period of the solution $L$ increases while moving along the branch. The branch asymptotically approaches the homoclinic bifurcation point labelled as $(1)$, when the period of the solution diverges. The limiting solitary wave solution $(1)$ corresponds to a depression region (a hole), as shown in the right panel of Fig.\,\ref{F2}(d). All wave profiles are shown in the frame of reference of the vertical solid plane and propagate from left to right. The wave $(1)$ is slower than the neutrally stable wave and is sometimes referred to as a negative solitary wave, or negative soliton. The period-two solutions obtained for $n=2$ undergo a branching point bifurcation (labelled as BP$_{(1,2)}$). At the first branching point BP$_1$, a new branch of the periodic solution is born, which continues towards larger values of $v$ and asymptotically approaches  the homoclinic point $(2)$. The corresponding solitary wave is a bulge (or a drop) that travels much faster than the neutral wave, as shown in Fig.\,\ref{F2}(c).  Such waves are referred to as positive solitons. At the branching point BP$_2$, the second branch of periodic waves is born, which asymptotically approaches the homoclinic point $(3)$. The corresponding negative solitary wave is a double-hole solution, as shown in the right panel of Fig.\,\ref{F2}(b). For simplicity, we restrict ourselves here to $n=1,2$ only. However, it should be emphasized that for every combination of the film thicknesses $d_i$ there exists an infinite countable set of positive and negative solitary waves, with each wave characterised by the number of humps (maxima or minima), similar to one-layer films \cite{pumir83}.

Crucial for the long-time dynamics of a nonlinear wave train is the observation that the homoclinic bifurcation occurs at a single point in the diagram in Fig.\,\ref{F2}(a), i.e. the solitary wave has a unique travelling speed $v$ and a unique profile with a certain amplitude $\text{amp}_1$.  This result will be generalized in Section\,\ref{homcon} to show that there exists a discrete set of homoclinic orbits characterised by multiple humps in the wave profile, each occurring at a particular value of the travelling speed. This feature is identical to the discrete spectrum of solitary waves found in one-layer films \cite{pumir83,CHANG96}. As a result, merging waves cannot grow indefinitely and coarsening becomes arrested after the waves have reached the largest possible amplitude. Such a growth-limiting mechanism was first formulated in connection with the rolling waves described by the shallow-water equation \cite{balmforth_mandre_2004} and then later observed experimentally by studying rolling waves in granular matter \cite{Razis14}. More specifically, we explain the arrested coarsening as follows. Assume that in an infinitely extended system with fixed average film thicknesses the instantaneous film profiles consist of several smaller waves and one large wave, whose profile is close to the homoclinic orbit (3) in Fig.\,\ref{F2}(a). For the larger wave to grow even further, some smaller waves must vanish, by transporting the corresponding amount of fluid into the large wave. Because the size of the smaller waves (and, therefore, the amount of fluid) is arbitrary, this process is only possible, if there exists a continuous spectrum of stable solitary waves possibly with different travelling speeds close to $v=8.1889$ (orbit (3) in Fig.\,\ref{F2}(a)). However, the results in this section and in Section\,\ref{homcon} show that there exists a discrete spectrum of solitary waves, each characterised by the number of humps and a certain value of $v$. In other words, another solitary wave with a speed arbitrarily very close to $v=8.1889$ simply does not exist.  Of course, the above considerations do not involve the notion of stability of the multiple-humps solitary waves. The merging of two waves is not possible if the resultant larger wave is unstable.

\begin{figure}[ht]
   \includegraphics[width=\columnwidth]{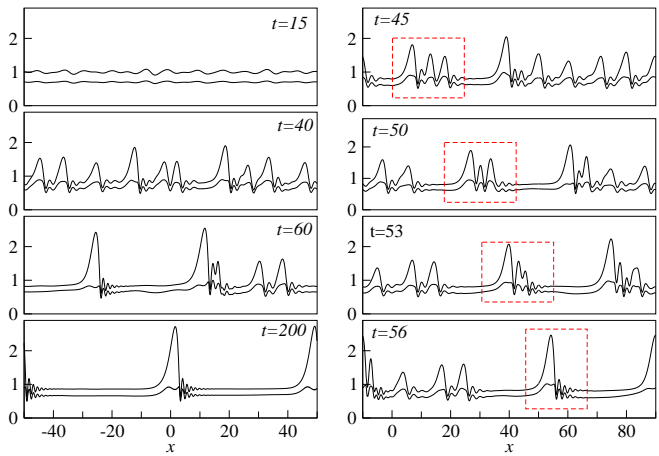}
   \caption{\label{F3} Temporal evolution of the initially flat two-layer film with $\sigma=\gamma_i=\mu=\rho=1$ and $d_2/d=0.7$. The left panel shows different stages of wave evolution. The right panel shows the dominating coarsening mode, whereby a faster travelling larger wave catches up and joins with a slower travelling smaller wave. }
   \end{figure}

   We use direct numerical simulations of the integrated-layer model Eqs.\,(\ref{eq1},\ref{eq2}) to gain an insight into the stability of nonlinear travelling waves, their interaction and coarsening behaviour. A comprehensive stability analysis based on the linearization of Eqs.\,(\ref{eq1},\ref{eq2}) about the travelling wave solutions falls beyond the scope of this study and will be discussed elsewhere. For the parameters as in Fig.\ref{F2}, we follow the temporal evolution of initially flat two-layer film in a large domain of size $100$ with periodic boundaries. The size of the domain is approximately 25 times larger than the wavelength of a neutrally stable wave $2\pi/k_c=3.9$.  Different stages of the temporal evolution are shown in the left panel in Fig.\ref{F3}. In the early stages, a zig-zag mode is formed in agreement with the linear stability analysis. The waves with a larger amplitude travel faster than smaller waves. This causes the waves to collide, leading to further amplitude growth and coarsening, i.e. a gradual transition from small-scale to large-scale deformations. The collision of waves eventually stops when several almost identical solitary waves are formed, each similar in shape to the positive solitary wave $(2)$ in Fig.\ref{F2}. A time-lapse of a single collision process is shown in the right panel in Fig.\ref{F3}. This scenario is similar to the so-called inelastic wave collision observed in one-layer liquid films \cite{Gollub94} and in rolling waves of granular matter \cite{Razis14}.

   As predicted by observing homoclinic bifurcations in Fig.\,\ref{F2}(a), the process of coarsening does not continue until a single solitary wave has survived. Instead, the growth of merging waves is arrested when the amplitude of the wave has reached its maximum possible value that, in turn is determined by the domain size and the average film thicknesses $d_i$. For parameters as in Fig.\ref{F3}, the waves reach the height of approximately $h_{\rm max}\approx 2.7$ at long times $t=200$. Further growth by means of wave merging is not possible because the resultant single wave would have the height larger than the maximum possible height of approximately $h_{\rm max}\approx 3.1$ (solitary wave (2) in Fig.\ref{F2}).

\begin{figure}[ht]
   \includegraphics[width=\columnwidth]{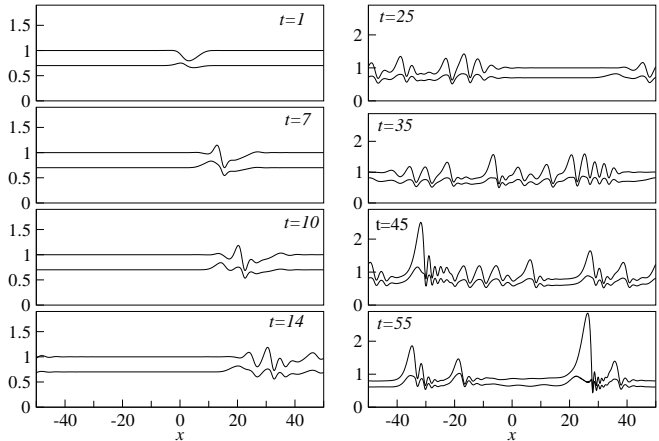}
   \caption{\label{F4}(a) Instability and break-up of the negative solitary wave.}
   \end{figure}

To investigate the stability of the negative solitary waves (propagating holes), such as solution $(1)$ in Fig.\ref{F2}(d), we create a single hole by applying a localized large-amplitude perturbation to the second layer, when the first layer is kept flat.  The temporal evolution of an isolated large hole is shown in Fig.\ref{F4}.  It is interesting to observe that the secondary smaller waves are formed in the front and tail of the hole, triggering its break-up and conversion into a nonlinear train of bulges, or drops. The subsequent collision between the drops leads to coarsening and eventually to a formation of a positive solitary wave similar in shape to $(2)$ in Fig.\ref{F2}.

\section{Mixed regime and mode competition}
\label{mixed}
\begin{figure}[ht]
   \includegraphics[width=\columnwidth]{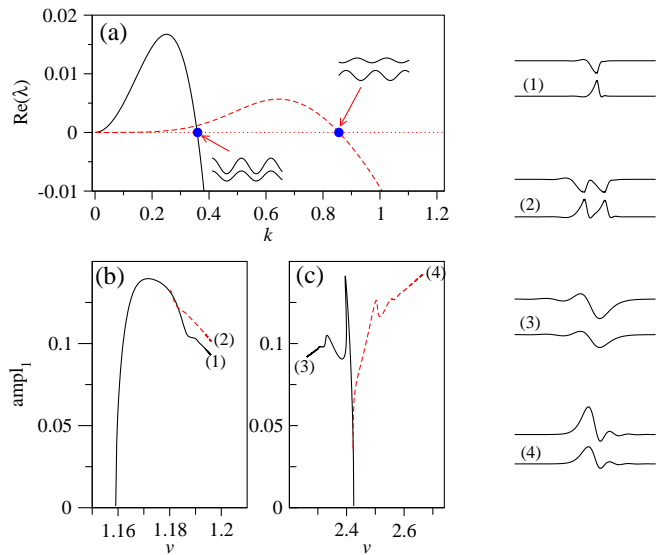}
   \caption{\label{F5} (a) Dispersion relations of the zig-zag and the varicose modes for $\mu=0.1$,  $\rho=2$, $\sigma=0.5$, $\gamma_2=10$, $\gamma_1=\sigma\gamma_2$ and $d_1/d=0.8$.  Critical $k_c$ for neutrally stable waves are marked by filled circles. (b,c) Bifurcation diagram of the periodic travelling waves that bifurcate from the two neutrally stable waves in (a).  Solid and dashed lines correspond to period-one and period-two solutions, respectively. Homoclinic bifurcation points are marked by (1,2,3,4). The corresponding solitary wave solutions are shown in the right panel.}
   \end{figure}

When the two-layer film is composed of non-identical fluids, the varicose and the zig-zag modes may be simultaneously unstable, as shown in Fig.\ref{F5}(a) for $\mu=0.1$,  $\rho=2$, $\sigma=0.5$, $\gamma_2=10$, $\gamma_1=\sigma\gamma_2$ and $d_1/d=0.8$. In this situation, we anticipate complex wave dynamics caused by the competition between the modes. Firstly, we construct a bifurcation diagram of the periodic travelling waves that bifurcate from the two neutrally stable waves, highlighted by filled circles in Fig.\ref{F5}(a). As in Fig.\ref{F2}, we only consider period-one and period-two solutions. The branches of the period-one solutions (solid lines) asymptotically approach the homoclinic bifurcation points marked by (1) for the varicose mode and by (3) for the zig-zag mode, as shown in Fig.\ref{F5}(b,c). The branches of the period-two solutions (dashed lines) bifurcate from the period-one solutions and asymptotically approach the homoclinic points (2) and (4). The fastest travelling solitary wave (4) is of the zig-zag type.

\begin{figure}[ht]
   \includegraphics[width=\columnwidth]{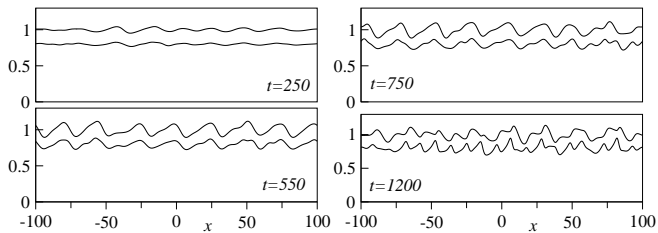}
   \caption{\label{F6} Temporal evolution of the initially flat two-layer film with parameters as in Fig.\,\ref{F5}.   }
   \end{figure}

Temporal evolution of the initially flat two-layer film with parameters as in Fig.\ref{F5} is shown in Fig.\ref{F6} in the domain of size $200$ with periodic boundaries. In agreement with Fig.\ref{F5}(a) the faster-growing zig-zag mode dominates the early stages of the evolution $(t<750)$. The varicose mode is shorter and travels at a slower speed. Eventually, after both modes have fully developed, the dynamics becomes rather chaotic $(t=1200)$. Similar to Fig.\ref{F3}, coarsening is arrested, when the waves reach their  maximum possible amplitude of approximately ampl$_1\approx 0.1$. It should be pointed out that an additional mechanism might be responsible for preventing the waves from further merging. Thus,
it is plausible to assume that coarsening is counterbalanced by the competition between a faster travelling zig-zag and a slower travelling varicose mode. In other words, the faster mode travels through the wave train of the slower mode which leads to a convective redistribution of the fluid. Such a "stirring" effect prevents the waves from colliding and thus, may lead to earlier arrest of coarsening. A qualitatively similar type of irregular dynamics in the early stages of the evolution was found in two coupled Kuramoto-Sivashinsky equations \cite{kliakhandler1999} that describe weakly nonlinear long wave evolution of a two-layer falling film in the presence of inertia. However, unlike in Fig.\ref{F6}, the simulations in Ref.\, \cite{kliakhandler1999} have been carried out for parameter values when only one mode (possibly zig-zag) is unstable.

\section{Homoclinic bifurcation of periodic travelling waves}
\label{homcon}
In this section, we study the homoclinic bifurcation of periodic travelling waves in more detail.  Thus, Eqs.\,(\ref{nonlin_eq3}) are written as a six-dimensional autonomous dynamical system
 \begin{eqnarray}
  \label{homcon_eq1}
  \partial_x h_1&=&p_1,~
   \partial_x p_1=p_2,~
   \partial_x p_2=\frac{r_1}{\gamma_1}-\frac{r_2}{\rho\gamma_1}\nonumber\\
    \partial_x h_2&=&s_1,~
   \partial_x s_1=s_2,~
   \partial_x s_2=-\frac{r_1}{\gamma_1}+\frac{r_2(\gamma_1+\gamma_2)}{\rho\gamma_1\gamma_2},\nonumber\\
   \end{eqnarray}
   with
   \begin{eqnarray}
  \label{homcon_eq2}
   r_1&=&-3+h_1^{-1}\left[-({\bm a}\cdot {\bm q})_1-v^2p_1+v({\bm {\partial q}}^{\sf T}  \cdot {\bm b}_1 \cdot {\bm q}) \right.\nonumber\\
   &+& \left. v({\bm {q}}^{\sf T}  \cdot {\bm b}_1 \cdot {\bm {\partial q}}) +p_1({\bm { q}}^{\sf T}  \cdot \partial_{h_1}{\bm b}_1 \cdot {\bm q})+s_1({\bm { q}}^{\sf T}  \cdot \partial_{h_2}{\bm b}_1 \cdot {\bm q})\right],\nonumber\\
   r_2&=&-3+h_2^{-1}\left[-\rho({\bm a}\cdot {\bm q})_2-v^2s_1+v({\bm {\partial q}}^{\sf T}  \cdot {\bm b}_2 \cdot {\bm q}) \right.\nonumber\\
   &+& \left. v({\bm {q}}^{\sf T}  \cdot {\bm b}_2 \cdot {\bm {\partial q}}) +p_1({\bm { q}}^{\sf T}  \cdot \partial_{h_1}{\bm b}_2 \cdot {\bm q})+s_1({\bm { q}}^{\sf T}  \cdot \partial_{h_2}{\bm b}_2 \cdot {\bm q})\right],\nonumber\\
  q_i&=&f_i+vh_i,~~{\bm {\partial q}}=(p_1,s_1).
\end{eqnarray}

Next, we relax the first two integral conditions in Eqs.\,(\ref{nonlin_eq4}) and choose the fluxes $f_i$ in such a way that Eqs.\,(\ref{homcon_eq1}) has a fixed point $h_i=d_i,~p_i=s_i=0$, $(i=1,2)$, with $d_i$ now replacing $f_i$ as parameters. Using the definition of fluxes in a flat two-layer film $q_i^{(0)}=f_i+vd_i$ we write
 \begin{eqnarray}
  \label{homcon_eq3}
  f_i=q_i^{(0)}-vd_i,
  \end{eqnarray}
where $q_i^{(0)}$ in dimensionless units are obtained from Eq.\,(\ref{stat_flux}) by replacing the factor $g\rho_1/(2\mu_1)$ with $3/2$.
Note that the above choice of $f_i$ is equivalent to the constant flux parametrisation used to study travelling waves in vertically falling one-layer liquid films \cite{pumir83,CHANG87,chang1993,CHANG96}.

\begin{figure}[ht]
   \includegraphics[width=\columnwidth]{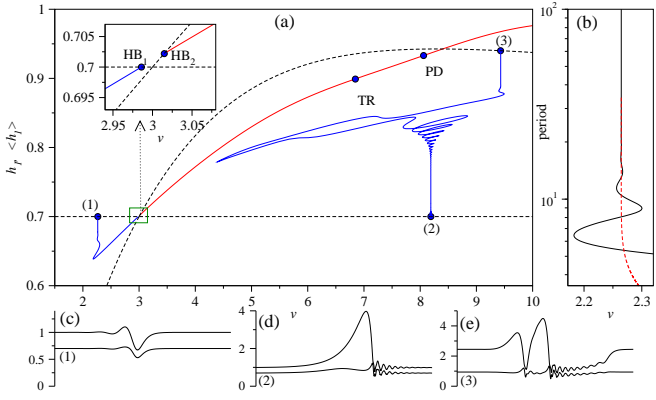}
   \caption{\label{F7} (a) Bifurcation diagram of the fixed points (dashed lines) and periodic solutions (solid lines) of Eqs.\,(\ref{homcon_eq1}) for parameters as in Fig.\,\ref{F2}. The inset shows the zoom of the region highlighted by the box.  (b) Period of the periodic orbits bifurcating from $\text{HB}_1$ (solid line), and the guideline $y=-\ln(v-v_c)$ to confirm logarithmic growth of the period as $v\rightarrow v_c=2.26442$ (dashed line).  (c,d,e) Homoclinic orbits found at points $(1)$, $(2)$ and $(3)$ in panel (a), respectively.}
   \end{figure}

We exemplify the bifurcation scenario of periodic solutions of system Eqs.\,(\ref{homcon_eq1}) using the parameters as in Fig.\ref{F2}, namely $\sigma=\gamma_i=\mu=\rho=1$ and $d_1=0.7,~d_2=0.3$.  In the first run, we follow the branch of fixed points of Eqs.\,(\ref{homcon_eq1}) using $v$ as a continuation parameter. Note that the bifurcation diagram of fixed points of Eqs.\,(\ref{homcon_eq1}) is rather complicated containing several connected branches, but for simplicity, we only show two branches which are relevant to the homoclinic orbits found earlier in Fig.\ref{F2}. In accordance with Fig.\ref{F2}, the fixed point $d_1=0.7, d_2=0.3$ has a Hopf bifurcation at $v=2.986$, as, labeled by $\text{HB}_1$ in the inset in Fig.\ref{F7}(a). The branch of periodic solutions born at $\text{HB}_1$ extends to $v<2.986$ and asymptotically approaches a homoclinic orbit (1), shown in Fig.\ref{F7}(c). The second branch of periodic solutions born at the second Hopf bifurcation point $\text{HB}_2$ extends to $v>3$ and contains a torus (TR) and a period-doubling (PD) bifurcation points. Additional branches of periodic solutions, bifurcating from the period-doubling points are not shown for simplicity. As a measure of periodic solutions, we use the average thickness of the first layer $\langle h_1\rangle = T^{-1}\int_0^T h_1(x)\,dx$, where $T$ represents the period of the solution.

The homoclinic bifurcation corresponds to a collision between a periodic solution and a fixed point with the period of the solution diverging. Following the branch that bifurcates from $\text{HB}_1$, we find the logarithmic divergence of the period, as shown in Fig.\ref{F7}(b). The dashed line in Fig.\ref{F7}(b) is a guideline $y=-\ln(v-v_c)$, where $v_c=2.26442$ is the value of $v$ where the homoclinic orbit is found. Note that solution (1) exactly coincides (up to a phase shift) with solution (1) in Fig.\ref{F2}(d). Remarkably, we find that the branch of periodic solutions approaches the homoclinic orbit non-monotonically, with every turning point of the branch representing a saddle-node bifurcation of the periodic solutions. This scenario is typical for Shilnikov bifurcation \cite{Shilnikov,Shilnikov2}, which predicts countably many saddle periodic orbits in the neighbourhood of a saddle-focus homoclinic loop if certain conditions on the eigenvalues of the corresponding fixed point are satisfied.

Next, we start continuation from the homoclinic orbit found in Fig.\ref{F2}(c) and find the branch of periodic solutions that extends between two homoclinic bifurcation points (2) and (3)  in Fig.\ref{F7}(a). Similar to point (1), the new branch is seen to approach the homoclinic solutions (2,3) non-monotonically. Interestingly, the new branch (2)-(3) is not born as a result of a Hopf bifurcation from a fixed point.


\begin{figure}[ht]
   \includegraphics[width=\columnwidth]{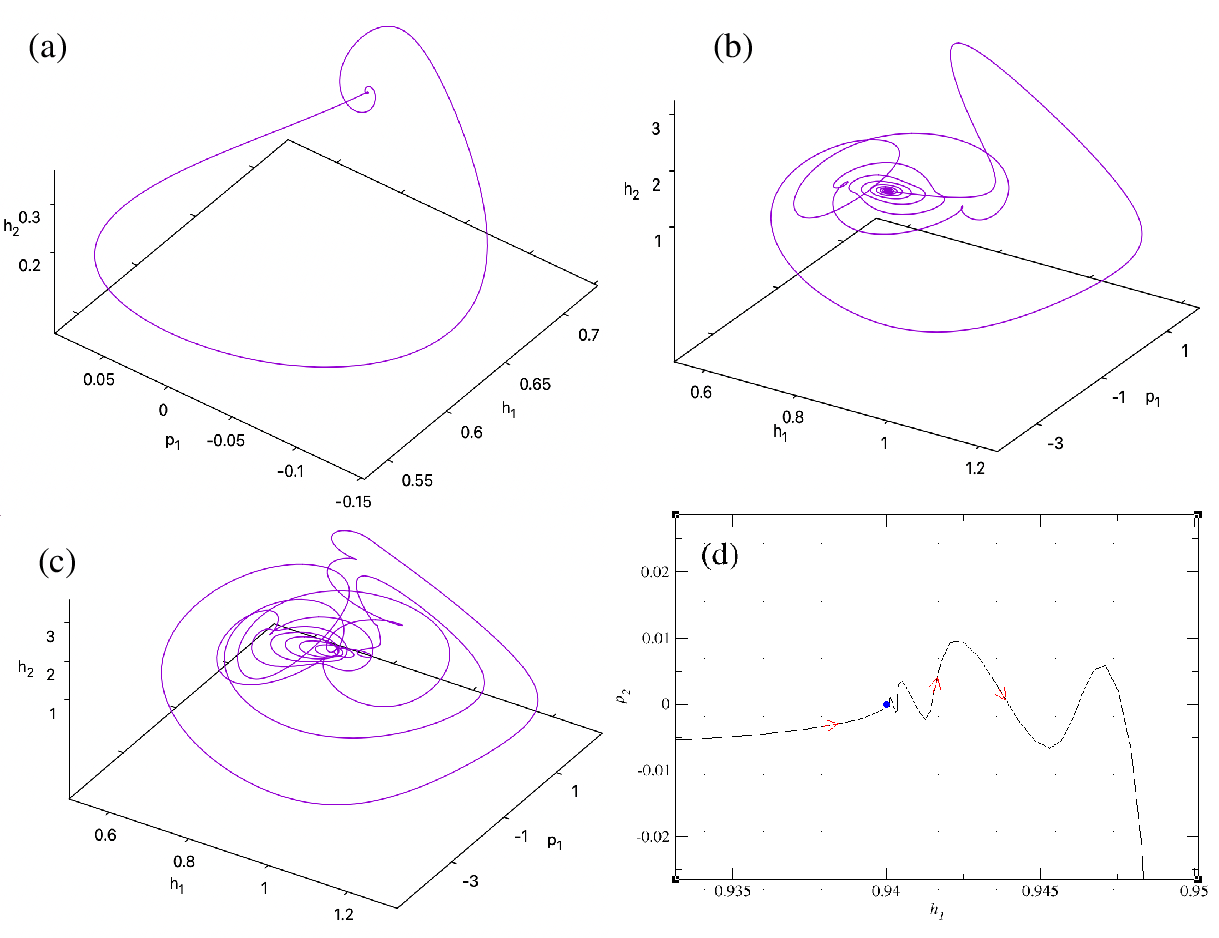}
   \caption{\label{F8} (a,b,c) Projection of the homoclinic orbits $(1)$, $(2)$ and $(3)$ from Fig.\ref{F7}(a), respectively onto three-dimensional phase space $(h_1,p_1,h_2)$. (d) Projection of orbit $(3)$ onto $(h_1,p_2)$ close to the fixed point $(h_1=0.94,p_2=0)$, highlighted by filled circle. Arrows indicate the directions of approach (departure) from the fixed point.}
   \end{figure}

Homoclinic orbits $(1,2,3)$ are shown in more detail in Fig.\ref{F8}(a,b,c), where we project them onto a three-dimensional phase space $(h_1,p_1,h_2)$ to highlight their saddle-focus character. Orbit $(3)$ is also of the saddle-focus type, as it can be seen from its projection onto the $(h_1,p_2)$ plane in Fig.~\ref{F8}(d), where arrows indicate the direction of approach (departure) from the fixed point $(h_1=0.94,p_2=0)$.
In order to make further connection with the Shilnikov theory of saddle-focus homoclinics \cite{Shilnikov2}, we compute the eigenvalues of the respective fixed points for orbits $(1,2,3)$. Thus, the fixed point $h_1=0.7,~h_2=0.3,p_i=s_i=0$ (orbit $(1)$) has the eigenvalues $\lambda=4.615624,~0.5018957 \pm 1.259518\,i,~-1.002640,~-2.308388\pm 4.706799\,i$. The leading eigenvalues are defined as the stable eigenvalue with the largest real part and the unstable eigenvalue with the smallest real part. Note that using the time-reversal transformation, ($x$ plays the role of "time" in Eqs.\,(\ref{homcon_eq1})) one can always reverse the flow of the dynamical system to change the sign of the eigenvalues, i.e. $\lambda\rightarrow -\lambda$. We observe that the leading eigenvalues of orbit $(1)$ are $\lambda_1=0.5018957 \pm 1.259518\,i$ (unstable) and $\lambda_2=-1.002640$ so that the so-called saddle quantity $\Sigma=\text{Re}(\lambda_1)+\text{Re}(\lambda_2)=0.5018957-1.002640<0$ is negative. These conditions exactly correspond to those of the Shilnikov theorem in higher dimensions \cite{Shilnikov2}. According to the theorem, there exist countably many periodic orbits in a neighbourhood of the saddle-focus homoclinic.

 Remarkably, however, the eigenvalues of the respective fixed points for orbits $(2,3)$ do not satisfy Shilnikov conditions. Indeed, for orbit $(2)$ the fixed point $h_1=0.7,~h_2=0.3,p_i=s_i=0$ at $v=8.1889$ has the eigenvalues $\lambda=3.414334,~0.2928146,~-0.4068085\pm 5.919676\,i,~-1.446766\pm 18.79867\,i$. The leading eigenvalues are $\lambda_1=0.2928146$ (unstable) and $\lambda_2=-0.4068085\pm 5.919676\,i$ (stable). Applying the time reversal transformation, the stability of the eigenvalues is exchanged, i.e. $\lambda_1=-0.2928146$ (stable) and $\lambda_2=0.4068085\pm 5.919676\,i$  (unstable) so that the saddle quantity $\Sigma=0.4068085-0.2928146>0$ is positive. This last condition violates Shilnikov's theorem condition \cite{Shilnikov2} and as the result, the statement of the theorem no longer applies.
  The sign of the saddle quantity $\Sigma$ is known to control the complexity of the dynamics in the neighbourhood of the homoclinic loop \cite{Shilnikov}. Thus, if $\Sigma$ is positive, the dynamics is known to be "simple", implying that there exists a stable periodic orbit, unlike for $\Sigma<0$, when the dynamics is chaotic. Of course, the stability of the periodic orbits of the dynamical system Eqs.\,(\ref{homcon_eq1}) has nothing to do with the spatio-temporal stability of the periodic travelling waves found for Eqs.\,(\ref{eq1}). The discussion of the phase flow properties of system Eqs.\,(\ref{homcon_eq1}) in the neighbourhood of the homoclinic orbits goes beyond the scope of this study.

  The eigenvalues of the fixed point $h_1=0.94,~h_2=1.50648,~p_i=s_i=0$ at $v=0.9432$ (orbit $(3)$) are $\lambda=0.9533101,~0.9328495\pm 1.008448\,i,~  -0.7198592 \pm 5.300151\,i,~-1.379291$. Here, the leading eigenvalues are $\lambda_1=0.9328495\pm 1.008448\,i$ (unstable) and $\lambda_2=-0.7198592 \pm 5.300151\,i$ (stable). The saddle quantity is positive, i.e. $\Sigma=0.9328495-0.7198592 >0$. Here again, the statement of the Shilnikov theorem doesn't apply.

  Having confirmed that orbit $(1)$ is born as a result of the Shilnikov saddle-focus homoclinic bifurcation, we leave the detailed investigation of the precise nature of the bifurcation points $(2,3)$ for later studies and move on to find homoclinic orbits that correspond to multiple humps in the wave profile. To this end, we use the idea of unfolding \cite{CHANG96} and extend the third and sixth equations in Eqs.\,(\ref{homcon_eq1}) by adding two dispersive terms
  \begin{eqnarray}
  \label{homcon_eq4}
  \partial_x p_2&=&\frac{r_1}{\gamma_1}-\frac{r_2}{\rho\gamma_1}+\delta p_2,\nonumber\\
   \partial_x s_2&=&-\frac{r_1}{\gamma_1}+\frac{r_2(\gamma_1+\gamma_2)}{\rho\gamma_1\gamma_2}+\delta s_2,
   \end{eqnarray}
   where $\delta$ represents the unfolding parameter.

\begin{figure}[ht]
   \includegraphics[width=\columnwidth]{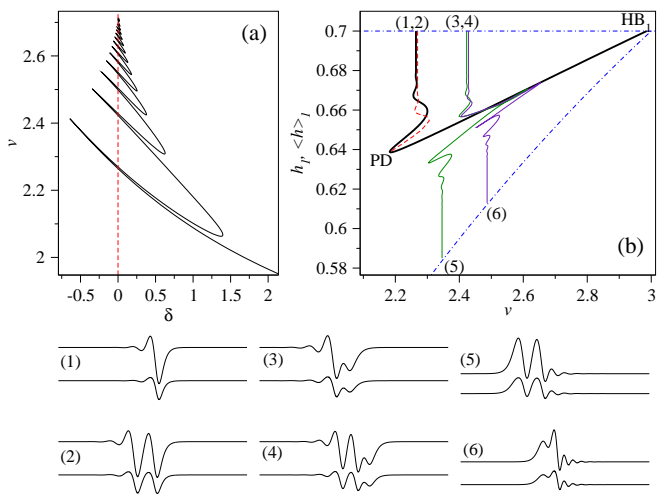}
   \caption{\label{F9}(a)  Unfolding of solitary wave solutions connected to the fixed point $h_1=0.7,~h_2=0.3$. Each intersection point of the solution branch with $\delta=0$ corresponds to a new multi-hump solitary wave. (b) Branches of periodic solutions connected to the first four solitary waves (1-4). The thick solid line corresponds to the primary periodic branch that bifurcates from the Hopf point $\text{HB}_1$ as in Fig.\ref{F7}(a). Dotted-dashed lines indicate fixed points of Eqs.\,(\ref{homcon_eq1}) with parameters as in Fig.\ref{F7}(a). Label PD indicates the period-doubling bifurcation point. Branches (3-5) and (4-6) are not connected to the primary periodic branch (thick solid line). The corresponding solitary wave profiles are shown in panels (1-6).}
   \end{figure}

 Starting with the solitary wave solution (1) in Fig.\ref{F7}(c), we fix the (large) period of the solution and continue it using parameter $\delta$ to obtain the branch of solitary waves in Fig.\ref{F9}(a). Each time the branch intersects the vertical dashed line $\delta=0$, a new multi-hump solitary wave solution is found. These new solutions can now be continued at fixed $\delta=0$ using $v$ as a principal continuation parameter to obtain the branches of periodic solutions, as shown in Fig.\ref{F9}(b) for the first four multi-hump solutions. For comparison, we also plot the primary period branch (thick solid line) that bifurcates from the Hopf point $\text{HB}_1$ in Fig.\ref{F7}(a). We find that the branch that bifurcates from solution (2) terminates at the period-doubling point PD on the primary periodic branch. The remaining multi-hump solutions give rise to periodic branches that are not connected with the primary branch. Thus we find that the solitary wave (3) is connected via periodic solutions with a solitary wave (5) and solitary wave (4) is connected with a solitary wave (6). Both (5,6) belong to the branch of fixed points (dotted-dashed lines) that are different from $h_1=0.7,~h_2=0.3$.
Similarly, the method of unfolding can be applied to find other multi-hump solutions starting from homoclinics (2,3) in Fig.\ref{F7}(a). For simplicity, these results are not shown and will be discussed in the future studies.

\section{Fluid transport modes and rupture of the second layer }
   \label{rupture}
  Next, we find a range of parameters that favours an intriguing novel dynamical regime, which only exists in two-layer films. Thus, when the density ratio $\rho$ is increased beyond $\rho=4$ with all other parameters fixed as in Fig.\ref{F5}, the developing varicose mode leads to the formation of a thin neck region in the second layer. With time, the thickness of the neck becomes vanishingly small and the second layer eventually ruptures. As the result, several isolated drops made of fluid $2$ may form on top of the first layer that transports these drops downstream as a conveyor belt. The corresponding temporal evolution of the initially flat film resembles that in Fig.\,\ref{F6} up until the moment of time when the neck of the varicose mode ruptures. At this stage, a true triple-phase contact line develops and the numerical simulations cannot be continued.

  To gain more insight into the varicose mode thinning and rupture, we compute the streamlines for period-one solutions in the coming frame of reference. Thus, we follow \cite{pot19} and construct two stream functions $\Psi_{(1,2)}(x,z)$ for each layer
  \begin{eqnarray}
  \label{stream}
  \Psi_1&=&({\bm m}\cdot {\bm q})_1\left(\frac{1}{2}z^3-z^2h_1\right)-\frac{1}{2}({\bm m}\cdot {\bm q})_2h_2z^2,\nonumber\\
  \Psi_2&=&\frac{1}{2}\mu({\bm m}\cdot {\bm q})_1\left( \frac{1}{3}(z-h_1)^3-h_2(z-h_1)^2\right)\nonumber\\
  &-&z[({\bm m}\cdot {\bm q})_1 h_1^2 +({\bm m}\cdot {\bm q})_2 h_1 h_2]\nonumber\\
  &+&\frac{1}{3}({\bm m}\cdot {\bm q})_1h_1^3+\frac{1}{2}({\bm m}\cdot {\bm q})_2h_1^2h_2,
  \end{eqnarray}
  with the fluxes $q_i$ in the laboratory frame of reference given by Eq.\,(\ref{nonlin_eq1}).
  The velocity fields $(u_i,w_i)$, $(i=1,2)$ in the laboratory frame of reference are given by $u_i=\partial_z \Psi_i$ and $w_i=-\partial_x \Psi_i$. In the frame of reference co-moving with the wave that travels at speed $v$, the stream functions must be transformed according to $\Psi_i \rightarrow \Psi_i -vz$.

\begin{figure}[ht]
\includegraphics[width=\columnwidth]{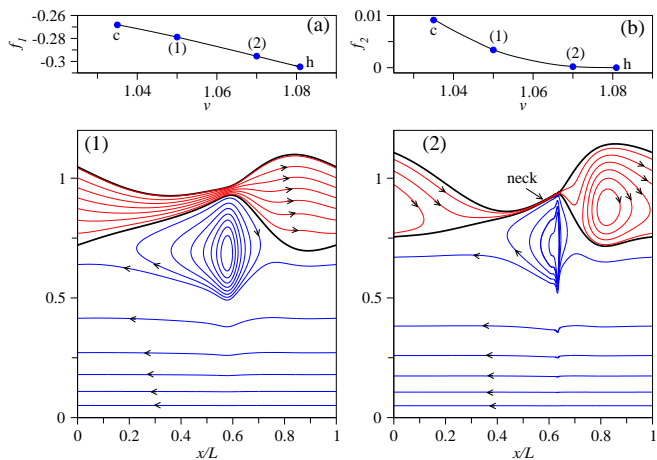}
\caption{\label{F10} (a,b) Fluxes $f_{(1,2)}$ in the co-moving reference frame for period-one solutions as a function of the travelling speed $v$ for $\mu=0.1$, $\rho=4$, $\sigma=0.5$, $\gamma_2=10$ and $d_2/d=0.8$.   Continuation starts at $v_c=1.035$ (point c), two selected points on the branch $(1,2)$ correspond to $v=1.05$ and $v=1.07$, respectively.  Panels (1,2): selected periodic solutions together with the streamlines in the co-moving frame of reference (waves travel from left to right).}
\end{figure}

  To obtain the branch of period-one solution with fixed average film thicknesses of the two layers, we proceed as in Figs.\,\ref{F2},\ref{F5} and continue the small-amplitude neutrally stable varicose mode using $v$ as a principal continuation parameter with three integral conditions Eqs.\,\ref{nonlin_eq4}. The fluid fluxes of period-one solutions in the co-moving frame of reference $f_i$ are shown in Fig.\,\ref{F10} (a,b). Continuation starts at $v=1.035$ (point c) and asymptotically approaches  the homoclinic bifurcation point (point h). Two selected points on the period-one branch correspond to $v=1.05$ and $v=1.07$ and are marked by labels (1,2), respectively. The corresponding solutions together with the streamlines, obtained from Eqs.\,(\ref{stream}), are shown in the panels (1,2).

   Two distinct fluid transport modes can be identified by studying the streamlines in the frame of reference co-moving with the wave as shown in Figs.\,\ref{F10}(1,2). In the first mode $(1)$, the fluid in the second layer overtakes the wave along the streamlines that have the same spatial period $L$ as the wave itself. The second mode $(2)$ is found in a varicose-type solution that has a thin neck and is characterised by the presence of a recirculation cell. The portion of the fluid trapped in the recirculation cells travels on average with the speed of the wave. Remarkably, as the period $L$ of the period-one solution increases, the neck becomes increasingly thinner, leading to an almost complete arrest of the flow in the second layer, while the flow in the first layer remains non-zero, as shown in Figs.\,\ref{F10} (a,b).  In this regime, an isolated rolling drop made of fluid $2$ is formed. The drop is transported downstream by the first layer which acts as a conveyor belt. Note that numerical simulation of Eqs.\,(\ref{eq1}) cannot be continued beyond the point of rupture of either of the two layers. Therefore, a stabilizing disjoining pressure must be added similar to \cite{pot05} to be able to reach the solution with a thin neck (2) in Fig.\,\ref{F10}.

\section{Conclusion}
\label{conc}
In conclusion, we have used a long-wave boundary-layer model of a two-layer liquid film with inertia to study nonlinear periodic and solitary two-dimensional rolling waves in a falling two-layer film. Linear stability of a flat film correctly predicts the existence of two types of surface and interface waves: the faster-travelling zig-zag mode and the slower-travelling varicose thinning mode.  The travelling speed of zero-amplitude neutrally stable waves that propagate downstream without growing or shrinking is shown to exactly coincide with the corresponding analytical expression, derived earlier from the full Navier-Stokes equation in the long-wave limit \cite{Kao65,kao1968}.

We formulate a boundary value problem to find periodic travelling waves with fixed average film thickness that bifurcate from a flat two-layer film. The efficient numerical continuation method \cite{krauskopf,dijkstra2014,doedel1981auto} is employed to follow the branch of periodic travelling waves starting from a small-amplitude neutrally stable wave, whose wavelength is determined from the linear stability analysis. Solitary waves are found as a result of the homoclinic bifurcation of periodic solutions, i.e. when the wavelength of the periodic wave diverges. As in falling one-layer films, we find two families of solitary waves: one family travels slower than the neutrally stable wave and is identified with depression regions or holes in otherwise flat layers. The second family travels faster than the neutrally stable wave with the wave profile resembling an elevation region, or a droplet preceded by a wavy front. We use the method of unfolding \cite{CHANG96} to find countably many solitary waves that travel at different speeds and are characterised by the number of "humps", i.e. maxima or minima in their profiles. Thus, a single "hump" wave corresponds to a single rolling two-dimensional drop or hole propagating downstream, while a multiple-hump wave may describe a cascade of multiple rolling drops or holes. Countably many solitary waves have also been found in falling one-layer films \cite{pumir83}.

By identifying the periodic travelling waves in two-layer films with limit cycles of a six-dimensional dynamical system, we demonstrate that there exist two distinct bifurcation mechanisms that lead to the birth of periodic solutions. On one hand, a limit cycle can be born from a fixed point as a result of a Hopf bifurcation. On the other hand, a limit cycle may undergo a homoclinic bifurcation with a diverging period, when it collides with a fixed point. Some of the found here homoclinics at a saddle-focus equilibrium have a negative saddle quantity, thus satisfying the Shilnikov condition \cite{Shilnikov,Shilnikov2} for countably many saddle periodic orbits in a neighbourhood of the homoclinic loop. However, not all homoclinics have a negative saddle quantity, implying that the dynamics in the six-dimensional phase space in the neighbourhood of the homoclinic orbit with a positive saddle quantity may be simple with trajectories converging to a stable limit cycle. Bifurcation analysis of periodic waves shows the existence of the solitary wave with the largest possible amplitude in the system with fixed average film thicknesses. As the result, a wave cannot grow above a certain limit, which prevents large waves from merging,  leading to the eventual arrest of coarsening in the system with fixed fluid volumes.

Direct numerical simulations of the model equations also shed light on the interaction of waves in the early stages of temporal evolution and the stability of the solitary waves. Similar to one-layer falling films \cite{Gollub94}, we find that in the early stage coarsening is dominated by inelastic wave collisions and merging. In agreement with previous studies, Refs.\,\cite{balmforth_mandre_2004,Razis14}, coarsening is arrested when the waves reach their largest possible amplitude. In the mixed dynamical regime, when both mode types are unstable, we find a highly irregular complex state with a faster zig-zag wave propagating through a train of slower varicose waves.

A distinct feature of the surface waves dynamics in two-layer falling films is the possibility of a new conveyor belt-like transport mode, when the second layer breaks down into isolated rolls supported by the layer adjacent to the solid plate, which transports these rolls downstream thus acting as a conveyor belt. We identified the pivotal role of the developing varicose mode in the scenario that leads to the formation of a think neck region in the second layer and its eventual rupture. It is plausible that this transport mode may also exist in two dimensions, allowing for an isolated drop made of the second fluid to be transported downstream by the first lubricant liquid layer.

Finally, we note that the behaviour of the solitary rolling waves investigated in this work is, in general, different from that of paradigmatic material solitary waves on shallow water surfaces that are described by the Korteweg-De Vries (KdV) equation \cite{Kor95} and from a large class of optical solitary waves observed in nonlinear optical waveguides \cite{Kivshar_book}, where the properties of light can be described using the nonlinear Schr{\"o}dinger equation \cite{Kivshar_book} or Maxwell's equations \cite{Goo92}. Indeed, for example, while both solitary rolling waves and KdV solitary waves of higher amplitude travel faster than those of lower amplitude, the effect of merging exhibited by the rolling solitary waves is different from the interactions between two KdV pulses that are known to pass one through without losing their identity \cite{Zab65}. A similar observation was made previously for single falling liquid films in the work \cite{Gollub94}, and it was also theoretically shown in \cite{Ker94} that in a physically relevant scenario of liquid films flowing along a vertical circular fibre there exists a critical film thickness above which solitary wave pulses can merge via inelastic-like interactions. While the behaviour of solitary waves in the two-layer film systems is even more complex than in single films, it is plausible that such systems could be used to predict and understand the physics of the break up of films into a periodic array of droplets, which is a phenomenon that is important in many technological processes \cite{Que90}.

\section{Acknowledgements}
 ISM was supported by the Australian Research Council through the Future Fellowship (FT180100343) program.
%

\end{document}